%% file: SIGIR2023.tex
\documentclass[sigconf]{acmart}

\AtBeginDocument{%
  \providecommand\BibTeX{{%
    \normalfont B\kern-0.5em{\scshape i\kern-0.25em b}\kern-0.8em\TeX}}}

\setcopyright{acmcopyright}
\copyrightyear{2023}
\acmYear{2023}
\acmDOI{XXXXXXX.XXXXXXX}

\acmConference[SIGIR '23]{The 46th International ACM SIGIR Conference on Research and Development in Information Retrieval}{July 23--27,
  2023}{Taipei, Taiwan}
\acmPrice{15.00}
\acmISBN{978-1-4503-XXXX-X/18/06}

\usepackage[inline]{enumitem}
\usepackage{graphicx}
\usepackage{color}
\usepackage{float}
\usepackage{subfig}
\usepackage[ruled]{algorithm2e}
\usepackage{algorithmic}
\usepackage{tabularx} 
\begin{document}

\title{$\rm \bf T^2Ranking$: A large-scale Chinese Benchmark for Passage Ranking}

\author{Xiaohui Xie}
\email{xiexiaohui@mail.tsinghua.edu.cn}
\affiliation{%
  \institution{DCST, Tsinghua University. Zhongguancun Lab.}
  \city{Beijing}
  \country{China}}
  
\author{Qian Dong}
\email{dq22@mails.tsinghua.edu.cn}
\affiliation{%
  \institution{DCST, Tsinghua University. Zhongguancun Lab.}
  \city{Beijing}
  \country{China}}

\author{Bingning Wang}
\email{bryantwwang@tencent.com}
\affiliation{%
  \institution{Tencent Inc.}
  \city{Beijing}
  \country{China}}
\author{Feiyang Lv}
\email{feiyanglv@tencent.com}
\affiliation{%
  \institution{Tencent Inc.}
  \city{Beijing}
  \country{China}}

\author{Ting Yao}
\email{tessieyao@tencent.com}
\affiliation{%
  \institution{Tencent Inc.}
  \city{Beijing}
  \country{China}}
\author{Weinan Gan}
\email{carrygan@tencent.com}
\affiliation{%
  \institution{Tencent Inc.}
  \city{Beijing}
  \country{China}}

\author{Zhijing Wu}
\email{wuzhijing.joyce@gmail.com}
\affiliation{%
  \institution{Beijing Institute of Technology}
  \city{Beijing}
  \country{China}}
\author{Xiangsheng Li}
\email{lixsh6@gmail.com}
\affiliation{%
  \institution{Tencent Inc.}
  \city{Beijing}
  \country{China}}

\author{Haitao Li}
\email{liht22@mails.tsinghua.edu.cn}
\affiliation{%
  \institution{DCST, Tsinghua University. Zhongguancun Lab.}
  \city{Beijing}
  \country{China}}

\author{Yiqun Liu}
\email{yiqunliu@tsinghua.edu.cn}
\affiliation{%
  \institution{DCST, Tsinghua University. Zhongguancun Lab.}
  \city{Beijing}
  \country{China}}
  
\author{Jin Ma}
\email{daniellwang@tencent.com}
\affiliation{%
  \institution{Tencent Inc.}
  \city{Beijing}
  \country{China}}
\renewcommand{\shortauthors}{Xie, et al.}

\begin{abstract}

Passage ranking involves two stages: passage retrieval and passage re-ranking, which are important and challenging topics for both academics and industries in the area of Information Retrieval (IR). 
However, the commonly-used datasets for passage ranking usually focus on the English language. 
For non-English scenarios, such as Chinese, the existing datasets are limited in terms of data scale, fine-grained relevance annotation and false negative issues.
To address this problem, we introduce $\rm T^2Ranking$, a large-scale Chinese benchmark for passage ranking.
$\rm T^2Ranking$ comprises more than 300K queries and over 2M unique passages from real-world search engines.
Expert annotators are recruited to provide 4-level graded relevance scores (fine-grained) for query-passage pairs instead of binary relevance judgments (coarse-grained). 
To ease the false negative issues, more passages with higher diversities are considered when performing relevance annotations, especially in the test set, to ensure a more accurate evaluation. 
Apart from the textual query and passage data, other auxiliary resources are also provided, such as query types and XML files of documents which passages are generated from, to facilitate further studies. 
To evaluate the dataset, commonly used ranking models are implemented and tested on $\rm T^2Ranking$ as baselines. 
The experimental results show that $\rm T^2Ranking$ is challenging and there is still scope for improvement. 
The full data~\footnote{The dataset is licensed under the \href{https://www.apache.org/licenses/LICENSE-2.0.html}{Apache License 2.0}} and all codes are available at \href{https://github.com/THUIR/T2Ranking/}{https://github.com/THUIR/T2Ranking/}

\end{abstract}



\maketitle

\input{Sections/1-introduction}

\input{Sections/2-related-work}

\input{Sections/3-task-definition}
\input{Sections/4-dataset-construction}

\input{Sections/5-data-statistics}
\input{Sections/6-experiment}
\input{Sections/7-conclusion}

\begin{acks}
This work is supported by the Tsinghua-Tencent Tiangong Institute for Intelligent Computing, the Beijing Academy of Artificial Intelligence~(BAAI) and the Quan Cheng Laboratory. 
\end{acks}

\bibliographystyle{ACM-Reference-Format}
\bibliography{sample-base}

\end{document}

%% file: Sections/1-introduction.tex

\section{Introduction}
\label{section:introduction}


Passage ranking is a crucial component of information retrieval systems.
The promising performance of passage ranking leads to satisfaction of search users and benefits multiple IR-related applications, e.g., question answering~\cite{aktolga2011passage} and reading comprehension~\cite{nishida2018retrieve}. 
Typically, passage ranking encapsulates two coherent stages, i.e., passage retrieval and passage re-ranking.
The goal of passage ranking is to compile a search result list ordered in terms of relevance to the query from a large passage collection. 
The first stage, passage retrieval, needs to recall relevant passages from a massive passage corpus. 
Hence, efficiency should also be considered besides effectiveness~\cite{fan2022pre}.
The second stage, passage re-ranking, may employ models that focus more on effectiveness to re-rank passages retrieved in the first stage.

To support the passage ranking research, various benchmark datasets are constructed.
Some of them support both the first-stage retrieval~(FR) and second-stage re-ranking~(SR) task, while others focus on the SR task.
We present the summary of data statistics of some common datasets in Table~\ref{table:summary_of_existing_datasets}.
Commonly-used datasets focus on English scenarios.
For example, Trec Complex Answer Retrieval~(Car)~\cite{dietz2017trec}, TriviaQA~\cite{joshi2017triviaqa} and MS-MARCO~\cite{nguyen2016ms}.
Among them, MS-MARCO is a large-scale dataset with 8.8 million passages. 
The queries are question-based, and human-generated answers are provided by annotators. 
Subsequently, a binary relevance score can be obtained by determining whether the answer related to the query exists in the passage; that is, relevant~(1) for passages containing the answer and non-relevant~(0) for those that don't. 
Following the success of MS-MARCO, similar datasets have also been constructed in the non-English community, such as Chinese.
For example, mMarco-Chinese~\cite{bonifacio2021mmarco} is the Chinese version of the original MS-MARCO with the help of machine translation.
$\rm DuReader_{retrieval}$~\cite{qiu2022dureader_retrieval} adopts a similar paradigm that generates binary relevance judgments for query-passage pairs from human-generated answers. 
Multi-CPR~\cite{long2022multi} is a multi-domain Chinese dataset for passage retrieval, with three different domains and a certain amount of human-annotated query-passage pairs.
Besides, Sogou-SRR~\cite{zhang2018relevance}, Sogou-QCL~\cite{zheng2018sogou} and TianGong-PDR~\cite{wu2020leveraging} are provided based on user logs from Sogou\footnote{https://www.sogou.com/}, a popular Chinese search engine.

\begin{table*}[h]
\caption{The data statistics of datasets commonly used in passage ranking. Qrys~(Psgs): Queries~(Passages). FR(SR): First~(Second)-stage of passage ranking, i.e., passage Retrieval~(Re-ranking). }
\label{table:summary_of_existing_datasets}
\begin{tabular}{cccccccc}
\toprule
Dataset            & Lang & \#Qrys  & \#Psgs  & Qrys.source    & Psgs.source       & Annotation   & Task   \\ \midrule
Trec Car~\cite{dietz2017trec}           & EN   & 2M   & 30M  & Wiki doc.   & Wiki doc.      & Binary       & SR     \\
TriviaQA~\cite{joshi2017triviaqa}           & EN   & 95K  & 650K & Trivia Web. & Wiki./Web doc. & Binary       & FR, SR \\
MS-MARCO~\cite{nguyen2016ms}           & EN   & 516K & 8.8M & User logs   & Web doc.       & Binary       & FR, SR \\ \midrule
Sogou-SRR~\cite{zhang2018relevance}         & CN   & 6K   & 63K  & User logs   & Web doc.       & Fine-grained & SR     \\
Sogou-QCL~\cite{zheng2018sogou}         & CN   & 537K & 9M   & User logs   & Web doc.       & Click labels & SR     \\
TianGong-PDR~\cite{wu2020leveraging}       & CN   & 70   & 11K  & User logs   & News doc       & Fine-grained & FR, SR \\
mMarco-Chinese~\cite{bonifacio2021mmarco}     & CN   & 516K & 8.8M & User logs   & Web doc.       & Binary       & FR, SR \\
Multi-CPR~\cite{long2022multi}     & CN   & 303K & 3M & User logs  & Result Title       & Binary       & FR, SR \\
$\rm DuReader_{retrieval}$~\cite{qiu2022dureader_retrieval} & CN   & 97K  & 8.9M & User logs   & Web doc.       & Binary       & FR, SR \\ \midrule
$\rm T^2Ranking$(Ours)    & CN   & 307K & 2.3M & User logs   & Web doc.       & Fine-grained & FR, SR \\ \bottomrule
\end{tabular}
\end{table*}

Although existing datasets facilitate the development of passage ranking applications, there are several limitations that need to be addressed:
\begin{enumerate}[label=(\arabic*),nosep]
\item The datasets are neither large-scale nor human-generated, especially in the Chinese community.
Sogou-SRR and Tiangong-PDR involve a limited number of queries. 
Although mMarco-Chinese and Sogou-QCL are large-scale, the former is based on translation while the latter only embraces click labels.
Recently, two passage-ranking datasets with considerable data scales are constructed, namely, $\rm DuReader_{retrieval}$ and Multi-CPR.
\item Fine-grained human annotations are limited.
Most datasets apply binary relevance annotations.
Since \citet{roitero2018fine} show the benefit of fine-grained relevance scales, recent work also investigates fine-grained relevance annotations beyond binary~(coarse) paradigms.
However, the number of fine-grained annotations is quite limited, for example, less than 100K in Sogou-SRR and TianGong-PDR.
\item False negative problem harms the accuracy of evaluation.
As \citet{arabzadeh2022shallow} point out, the existing passage ranking datasets suffer from the false negative problem, i.e., relevant results are labeled as irrelevant. 
This problem exists mainly due to limited human annotations in the large-scale dataset, which will harm the accuracy of the evaluation.
For example, for each query in Multi-CPR, only one passage will be marked as positive while others are regarded as negative.
A recent dataset, i.e., $\rm DuReader_{retrieval}$, attempts to ease this issue by asking annotators to manually check and relabel the passages in the top retrieved results pooled from multiple retrievers.
\end{enumerate}

In order to ensure high-quality training and evaluation of passage ranking models, we construct and release a new Chinese dataset, named $\rm T^2Ranking$, comprising of more than 307K question-based queries and over 2.3M passages extracted from 1.8M web documents. 
Specifically, we sample search queries from user logs of the Sogou search engine, a popular search system in China, and perform query preprocessing, such as filtering pornographic queries and non-interrogative queries, and removing similar queries, to obtain a clean and high-quality query set with 307K queries~(50K for the test set). 
For each query, we extract the content of corresponding documents from different search engines and remove vertical results~(e.g., image search results and video search results) and duplicate results for the following process.
To ensure the semantic integrity of each passage, we train and use a passage segment model to access passages from each document, which gives us around 1.3 passages per document. 
We then use a passage clustering approach to discard highly similar passages and generate the query-passage pool. 
Moreover, we also record query types and other auxiliary resources of documents to facilitate extending studies~(e.g., multi-modal tasks and out-of-domain~(OOD) tasks). 
For a given query and its corresponding passages, we hire expert annotators to provide 4-level relevance judgments of each query-passage pair and adopt an active learning-based data sampling to improve the efficiency and quality of annotation.
All hired annotators are full-time staff engaged in annotation work.

We carry out comprehensive analyses and present comprehensive statistics of the proposed dataset. 
Additionally, we conduct comprehensive experiments to evaluate the performance of multiple passage retrieval models as well as passage re-ranking models, on $\rm T^2Ranking$. 
The experimental results show that $\rm T^2Ranking$ is a highly challenging task and there is still potential for further performance improvement.

In summary, we make the following contributions:
\begin{itemize}[leftmargin=*,nosep]
  \item We build a large-scale Chinese dataset, named $\rm T^2Ranking$ for passage ranking~(retrieval and re-ranking). 
  $\rm T^2Ranking$ contains more than 300K queries and over 2M unique passages, and also comes with fine-grained relevance annotations, along with query types, document titles and XML files as multimodal information.
  \item We leverage multiple strategies to ensure the high quality of our dataset, such as using a passage segment model and a passage clustering model to enhance the semantic integrity and diversity of passages and employing active learning for annotation method to improve the efficiency and quality of data annotation.
  \item We conduct extensive experiments to evaluate the performance of existing passage retrieval and re-ranking models on $\rm T^2Ranking$.
  Experimental results show room for further improvement which might be brought by more sophisticated models in the future.
\end{itemize}


%% file: Sections/2-related-work.tex

\section{Related Work}
\label{section:relatedwork}


There are several benchmark datasets developed for passage ranking. 
For datasets that have relevance annotations for all query-passage pairs, both passage retrieval and passage re-ranking tasks can be tested. 
Other datasets, however, only focus on passage re-ranking tasks, providing relevance annotations only for query-passage pairs in which the passages have been extracted from the initial result lists recalled by the first-stage retrievers.
We use FR to denote the first stage of passage ranking, i.e., passage retrieval and SR to denote the second stage of passage ranking, i.e., passage re-ranking as shown in Table~\ref{table:summary_of_existing_datasets}.

Commonly used datasets for passage ranking are constructed for the English community.
Trec Complex Answer Retrieval~(CAR)~\cite{dietz2017trec} uses topics, outlines, and paragraphs extracted from Wikipedia. For the training set, a passage is considered relevant if it is found within the Wikipedia pages of the topic and non-relevant otherwise. The test set, comprised of 113 complex topics, has 50 passages per topic that are manually annotated.
TriviaQA~\cite{joshi2017triviaqa} gathers question-answer pairs from 14 trivia and quiz-league websites and passages from Wikipedia and web documents. 
MS-MARCO~\cite{nguyen2016ms} is widely utilized due to its large scale.
Unlike Trec Car and TriviaQA, queries in MS-MARCO are sourced from user-generated queries, which are question-based, from the Bing search engine\footnote{\url{https://www.bing.com}}.
The Passages are extracted from realistic web documents returned by the same search engine.
Then human editors are recruited and instructed to create a natural language answer with the correct information extracted strictly from the passages provided given particular queries.
The relevance levels of passages in both TriviaQA and MS-MARCO are determined in a binary fashion, based on whether or not the passages contain facets of the true answer to a given query.

For the Chinese community, there exist several datasets designed for training and evaluating passage ranking models. 
Drawing upon the Sogou search engine, three datasets have been established, namely Sogou-SRR~\cite{zhang2018relevance}, Sogou-QCL~\cite{zheng2018sogou} and TianGong-PDR~\cite{wu2020leveraging}.
Sogou-SRR~(Search Result Relevance) consists of 6K queries and corresponding top 10 search results.
For each search result, the screenshot, title, snippet, HTML source code, parse tree, URL as well as a 4-grade relevance score and the result type are provided.
Sogou-QCL is a large-scale dataset compromised of 537K queries and more than 9 million Chinese web pages.
Rather than human-generated relevance judgments, relevance levels of query-result pairs are assessed based on click labels.
Queries from Tiangong-PDR are collected from Sogou's search logs, while passages are obtained from Web pages data from the Sina news website\footnote{\url{https://www.sina.com.cn/}}.
Moreover, four-grade human-assessed relevance labels for each query-passage pair are available.
Besides, mMarco-Chinese~\cite{bonifacio2021mmarco} is constructed via machine translation from MS-MARCO. 
However, these datasets are not large-scale and/or human-generated. 
Recently, \citet{qiu2022dureader_retrieval} propose a new dataset, named $\rm DuReader_{retrieval}$, for benchmarking the passage retrieval models from Baidu search\footnote{\url{https://www.baidu.com/}}.
Similar to MS-MARCO, queries in $\rm DuReader_{retrieval}$ are question-based, and human-generated answers are collected to access the relevance levels of passages. 
\citet{long2022multi} build Multi-CPR which is a multi-domain dataset for passage ranking. 
Queries and passages for Multi-CPR are gathered from three different vertical search systems: E-commerce, Entertainment Video, and Medical. 
Rather than being extracted from web documents, passages in Multi-CPR refer to titles of search results, such as product titles in E-commerce search, resulting in shorter passage lengths. Human annotators have been recruited to judge the relevance level (binary) of the query-passage pairs. For each query, the most semantically relevant passage is marked as positive, while the others are marked as negative.

%% file: Sections/3-task-definition.tex

\section{Task Definition}
\label{section:taskdefinition}

In this section, we formally define the tasks in $\rm T^2Ranking$. Our proposed dataset focuses on two stages of passage ranking, namely, passage retrieval and re-ranking. This aligns with the pipeline of modern information retrieval systems, which follows the retrieval-then-re-ranking paradigm.

The goal of passage retrieval is to retrieve candidate passages in response to a given query.
Given a query $q$, a retrieval model is used to retrieve a candidate set of passages $\mathcal{K}=\{\mathbf{p}_j^{\mathbf{q}}\}_{j=1}^{K}$ from a large corpus $\mathcal{G}=\{\mathbf{p}_i\}_{i=1}^{G}$, where $K\ll G$.
In particular, a passage consists of a sequence of words $\mathbf{p}=\{w_p\}_{p=1}^{|\mathbf{p}|}$, where $|\mathbf{p}|$ represents the length of passage $\mathbf{p}$.
Similarly, a query is a sequence of words $\mathbf{q}=\{w_q\}_{q=1}^{|\mathbf{q}|}$.
The main challenge in passage retrieval lies in efficiently retrieving the relevant passages for a query, given the vast number of passages in the corpus.
Following retrieval, re-ranking is proposed to derive a permutation over $\mathcal{K}$, such that the more relevant passages are ranked higher in the list. 
In contrast to the retrieval task, the re-ranking task demands that models have a strong capability for relevance modeling, which is capable of capturing subtle semantic differences between relevant passages in the candidate set $\mathcal{K}$.


%% file: Sections/4-dataset-construction.tex
    
\section{Dataset Construction}
\label{section:datasetconstruction}

In this section, we present the construction details of $\rm T^2Ranking$. We begin by introducing the overall pipeline of dataset construction, which includes query sampling, passage extraction, and relevance annotation. 
We then provide important technical details used in the data construction, such as model-based passage segmentation, clustering-based passage de-duplication, and active learning-based data sampling.

\subsection{Overall Pipeline}

The overall pipeline of constructing $\rm T^2Ranking$ involves several steps, including query sampling, document retrieval, passage extraction and fine-grained relevance annotation.

\noindent \textbf{Query sampling.} 
We sample real user queries from the query pool of Sogou and perform pre-processing (e.g. de-duplication and normalization of redundant spaces and question marks) to obtain a clean query dataset. Then, we filter out pornographic, non-interrogative and resource-request type queries and queries that might include user information from $\rm T^2Ranking$ using an intent analysis algorithm, to ensure that the dataset consists only of high-quality, question-based queries.
Note that resource-request-type queries are used to search for specific music, film resources, etc.

\noindent \textbf{Document retrieval.} 
We retrieve a comprehensive set of documents for each query from popular search engines such as Sogou, Baidu, and Google, taking advantage of their vast resources and expertise in indexing and ranking web content. 
This helps to reduce the issue of false negatives, as each system covers different parts of the web and can return different relevant documents, hence improving the overall coverage of our dataset. 

\noindent \textbf{Passage extraction.} 
The construction of passages in $\rm T^2Ranking$ involves segmentation and de-duplication.
Rather than using a heuristic approach to segment passages from a given document~(e.g, conventionally determining the start or the end of passages by line breaks), we employ a model-based method for passage segmentation to maximize the preservation of complete semantics in each passage~(detailed in Section~\ref{subsection:modelbasedpassagesegementation}).
Additionally, we introduce a clustering-based technique to enhance the efficiency of annotation and maintain the diversity of the annotated query-passage pairs~(detailed in Section~\ref{subsection:clusteringbasedpassagededuplication}). 
This approach effectively removes nearly identical passages that are retrieved by a particular query. 
The resulting segmented and de-duplicated passages are subsequently merged into the passage collection for $\rm T^2Ranking$.

\noindent \textbf{Fine-grained relevance annotation.} 
All hired annotators are experts in providing annotation for search-related tasks and have engaged in labeling work for a long time.
At least three annotators provide 4-level fine-grained annotations for each query-passage pair. 
Specifically, if the annotations are inconsistent among the first three annotators for a particular pair~(three annotators provide three different scores), a fourth annotator will be asked to access it. 
In cases where all four annotators are inconsistent, the query-passage pair is considered to be too ambiguous to determine the required information and will be excluded from the dataset. 
The final relevance label for each query-passage pair is determined by major voting.
Following the criteria of TREC benchmarks~\cite{craswell2021trec}, we also define the instructions of 4-level relevance annotation as:
\begin{itemize}[nosep]
    \item \textbf{Level-0.} There is a complete mismatch between the content of the query and the passage.
    \item \textbf{Level-1.} The passage is relevant to the query, but it does not meet the required information needs of this query.
    \item \textbf{Level-2.} The passage is relevant to the query and partly satisfies its information needs.
    \item \textbf{Level-3.} The passage content is customized to satisfy the information needs of the query and precisely contains the answer to the query.
\end{itemize}

We show several examples in Table~\ref{tab:example}.
The fine-grained 4-level annotation enables accurate evaluation of passage re-ranking tasks.
Notably, for the retrieval task, we consider \textbf{Level-2} and \textbf{Level-3} passages as relevant passages, and all other passages are regarded as irrelevant passages. 
\begin{table*}[]
\caption{Examples for annotation of query-passage pair.}
\label{tab:example}
\begin{tabularx}{\textwidth}{l|X|X|X}
	\toprule
\#Annotation  & Query & Passage & Explanation \\ \midrule
0 &Does burning tea leaves with chrysanthemums produce tea polyphenols?      &   Tea and chrysanthemum are allowed to be infused together. Chrysanthemum itself has the effect of...      &  There is no mention of query in the passage and it does not meet the information needs of query at all.           \\ \midrule
1 & What does cervical cancer stage IIB mean?      &  Stage IB cervical cancer means that the cancer is confined to the cervix and the colposcopy reveals lesions larger...       &  This passage pertains to cervical cancer, but it is not suitable for the given query, which requests information on stage IIB.    \\ \midrule
2 & What causes excessive sweating in children? &    A child's sweating in bed is due to night sweats in children...     &   The query is not limited to the topic of sweating during sleep only.          \\ \midrule
3 &   What flavour bait does the chub like?    &  The best bait for chub is a feed with an aromatic, fishy or smelly smell and fresh...       &   The answer is precisely contained in the passage.    \\ \bottomrule
\end{tabularx}
\end{table*}

Notably, when processing the test queries, we utilize the strategy of annotating all query-passage pairs after the passage segmentation process, which attempts to mitigate the problem of false negatives in our test set and hence provides a more precise evaluation of the retrieval and re-ranking performance.
For the training queries, we employ the aforementioned clustering-based method to du-duplicate the passages which are then presented to the recruited expert annotators to obtain 4-level fine-grained annotations. 
This strategy not only enhances the efficiency of annotation but also maintains diversity in the annotated query-passage pairs.
Besides, the success of the active learning strategy motivates us to rich the information involved in our training samples by choosing informative query-passage pairs for annotation.
The key idea behind active learning is that by allowing the model to select which training samples it wants to learn from and focus on samples that are most valuable for improving its performance, leading to more efficient annotation.
In $\rm T^2Ranking$, we design an active learning-based method to annotate the training data in an iterative manner~(detailed in Section~\ref{subsection:activelearningbaseddatasampling}).
Overall, the data construction pipeline of $\rm T^2Ranking$ is formally defined in Alg.~\ref{alg:construction}.
\input{Sections/4-algorithm4construction.tex}

\subsection{Model-based Passage Segmentation}
\label{subsection:modelbasedpassagesegementation}
Typically, in existing datasets, the passages are segmented from documents according to a natural paragraph or sliding window with a fixed length. However, the natural paragraph-based segmentation usually results in an excessively long passage containing multiple topics, considering most web documents are not well-written.
Besides, the sliding window-based segmentation often leads to a lack of complete semantics in a passage~\cite{qiu2022dureader_retrieval, cheng2023layout}, thereby reducing the reliability of the dataset for the evaluation of the passage retrieval and re-ranking. 

\begin{figure}
    \centering
    \includegraphics[width=\linewidth]{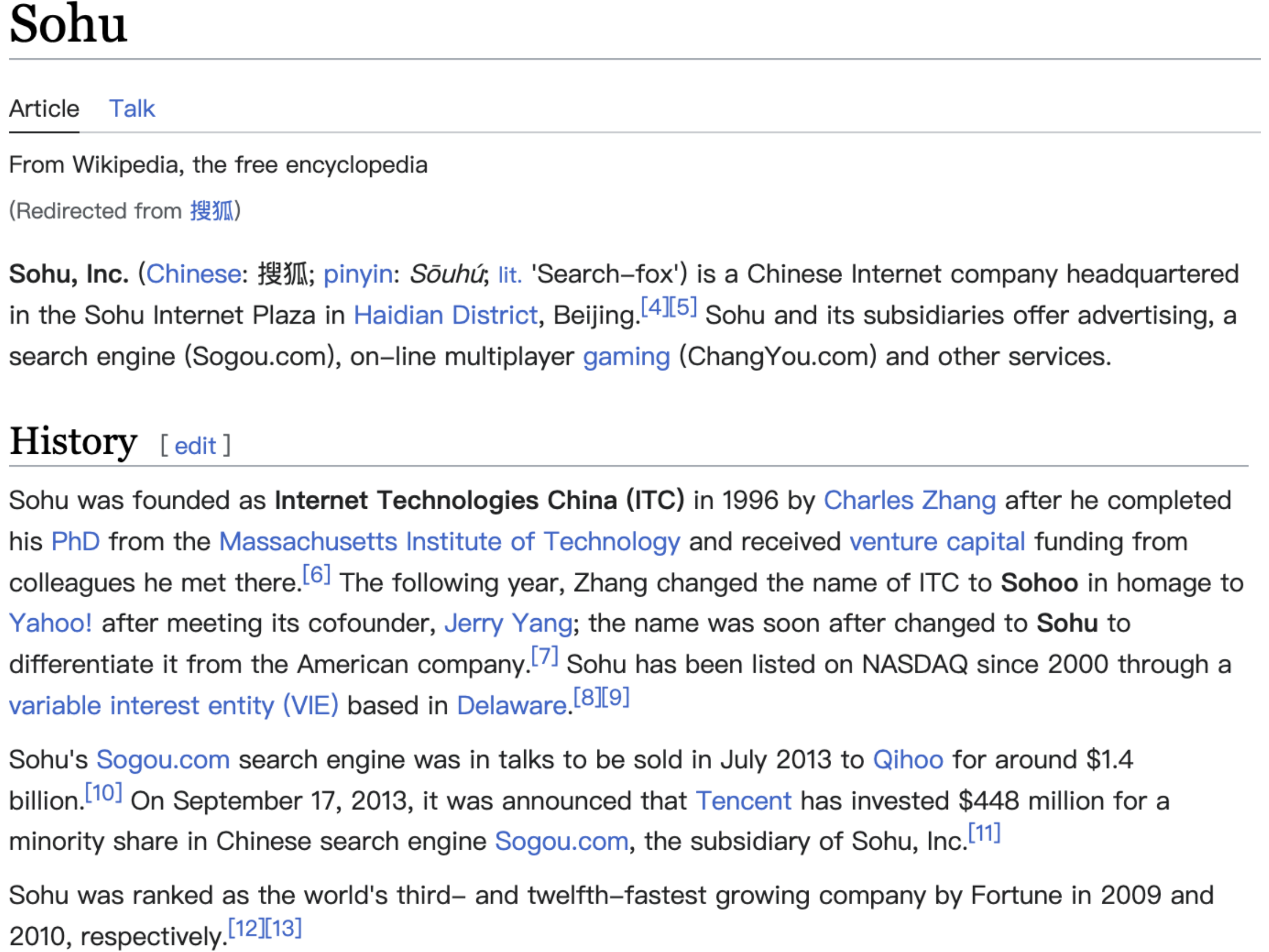}
    \caption{Illustration for a web document from Wikipedia which is well-written with clearly defined paragraphs.}
    \label{fig:baike}
\end{figure}

To address this issue, we propose a model-based method for passage segmentation. 
A segmentation model is trained on well-written web documents using the sequence labeling task. 
Specifically, we use the Sogou Baike\footnote{https://baike.sogou.com/}, Baidu Baike\footnote{https://baike.baidu.com/} and Chinese Wikipedia\footnote{https://zh.wikipedia.org/} as the training data, given that these web documents are generally well-written and their natural paragraphs are clearly defined. 
An example English version Wikipedia is shown in Figure~\ref{fig:baike}.
Given a web document $\mathbf{d}=\{w_d\}_{d=1}^{|\mathbf{d}|}$, we utilize a segmentation model $Seg(\cdot)$ to determine whether a given word $w_d$ should be separated.
Formally, the sequence labeling task can be defined as
\begin{equation}
    \hat{y}_d = Seg(\mathbf{d})_{w_d},
\end{equation}
\begin{equation}
    \mathbf{L}_{s} = \text{CrossEntropy}(y_d, \hat{y}_d),
\label{eq:seqloss}
\end{equation}
where the $\hat{y}_d$ is the predicted score for segmentation. The true label $y_d$ represents whether the word $w_d$ is the last word of a paragraph. 
The segmentation model $Seq(\cdot)$ is trained based on the loss defined in Eq.~\ref{eq:seqloss}.
If $\hat{y}_d\geq \sigma$, then the passage is segmented from its document by the $d$-th word. $\sigma$ is a hyperparameter that controls the degree of segmentation. The smaller the value of $\sigma$, the more passages are segmented.

\subsection{Clustering-based Passage De-duplication}
\label{subsection:clusteringbasedpassagededuplication}
Annotating a large number of highly similar passages on the web would be redundant and meaningless. 
In this paper, we propose a clustering-based method for passage de-duplication, which leads to more efficient annotation. 
Specifically, we employ a hierarchical clustering algorithm, Ward~\cite{ward1963hierarchical}, to unsupervisedly cluster similar passages together.
The passages in the same cluster are considered nearly duplicated.
Consequently, we select only one passage from each cluster for annotation.
It is worth noting that we only conduct the de-duplication in the training set. 
For the queries in the test set, we annotate all the passages obtained from the passage segmentation model to alleviate the false negative issue as much as possible. 
Intuitively, passages that are nearly identical under a specific query provide little information gain to a ranking model compared to passages with significant differences. 
Practically, false negatives within the same cluster as an annotated true positive can be easily filtered by a cross-encoder~\cite{qu2020rocketqa}.
Therefore, we employ the de-duplication to save the annotation cost while retaining more diverse training samples for improving model performance.

\subsection{Active Learning-based Data Sampling}
\label{subsection:activelearningbaseddatasampling}
In practice, we observe that not all training samples can further enhance the ranking model's performance. Training samples, that can be easily predicted accurately by a model, are unlikely to provide useful information for model training.

\begin{figure}
    \centering
    \includegraphics[width=\linewidth]{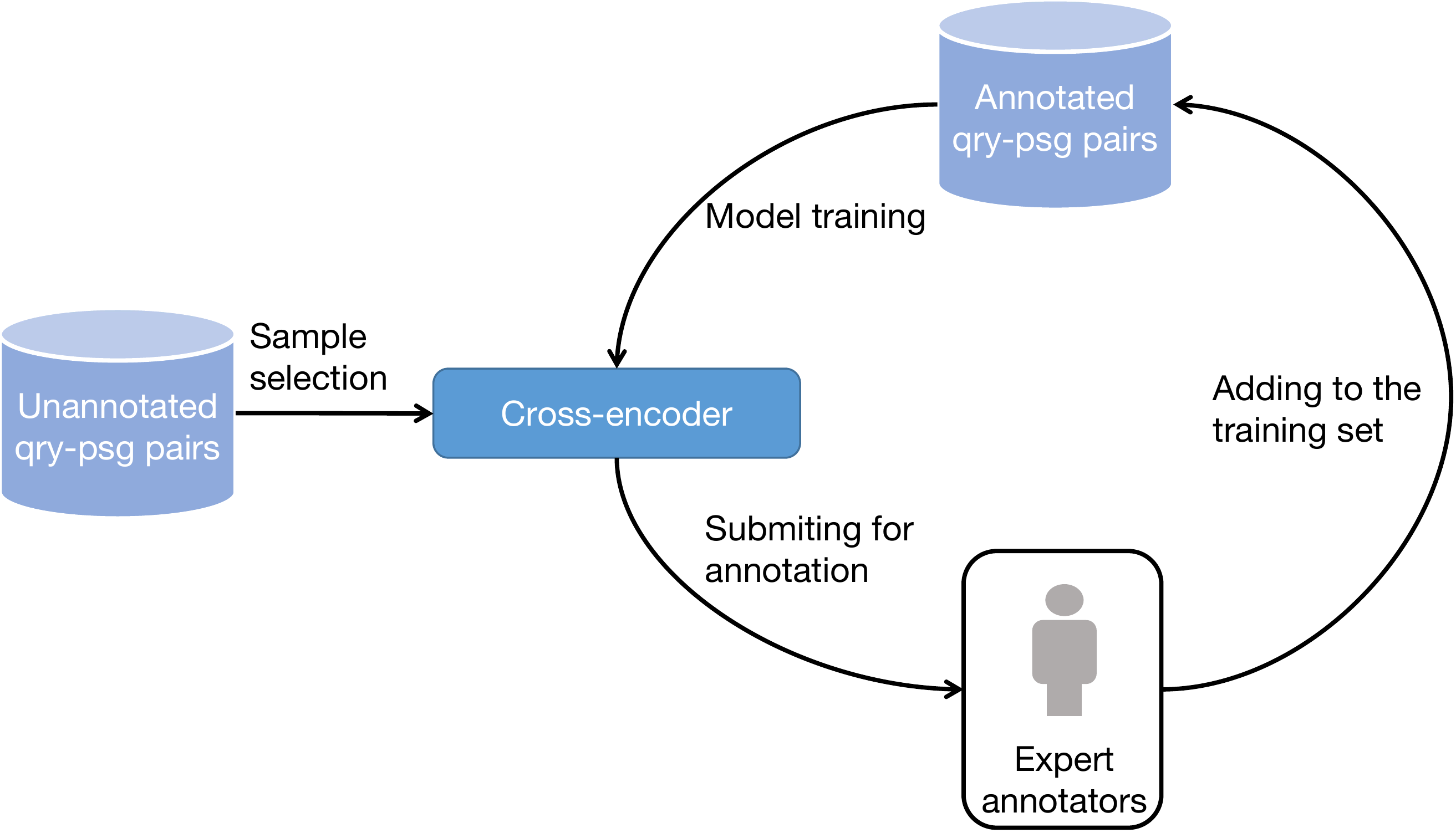}
    \caption{Illustration for the framework of active learning.}
    \label{fig:active}
\end{figure}

To address this issue, we borrow the light of active learning~\cite{ren2021survey}, using a model to choose more informative training samples for further annotations.
Active learning is a framework that enables models to participate in the data annotation process. The aim of active learning is to minimize the amount of annotated data required while maintaining or improving model performance. Formally, active learning is an iterative process where the model makes predictions on a pool of unannotated samples. The samples with the highest uncertainty or informativeness are selected for annotation by annotators, and the annotated samples are added to the training data. The model is then updated with the newly annotated data.
The framework of active learning is illustrated in Figure~\ref{fig:active}.
Concretely, a query-passage re-ranking model, specifically a cross-encoder, is trained using data constructed from the initial stage. In the second stage, unannotated query-passage pairs are obtained and evaluated for relevance by the cross-encoder. Pairs with high confidence scores are filtered out as they do not provide significant information for further performance improvement, while pairs with low confidence scores, which are typically considered noise samples, are also eliminated.
The remaining pairs are submitted to annotators for fine-grained annotation. The annotated query-passage pairs are then added to the training set and the cross-encoder is updated with newly acquired samples.

%% file: Sections/4-algorithm4construction.tex
\begin{algorithm}[!tb]
\caption{The pipeline of dataset construction.}
\label{alg:construction}
\KwIn{Query pool $\mathcal{Q}$, document pool $\mathcal{D}$, passage segmentation model $Seg(\cdot)$, cross-encoder $CE(\cdot)$ and expert annotator $\mathcal{H}$}
\KwOut{Fine-grained relevance labels $\mathcal{L}$}
\hrulefill 


\FuncSty{${\mathrm{DatasetConstruction}}(\mathcal{Q}, \mathcal{D}, \mathcal{H})$} \Begin{
    $\mathbf{Q} = \text{Sample}(\mathcal{Q})$; \% sampling a set of queries $\mathbf{Q}$;\\
    $\mathcal{L}=\emptyset, \mathcal{P}=\emptyset;$ \% initialising a label set and a passage set; \\
    
    \For{$\mathbf{q} \in \mathbf{Q}$}{
        \%retrieving a set of documents $\mathbf{D}$ for query $q$ via multiple search engines;\\
        $\mathbf{D} = \text{MultiSearchEngines}(\mathbf{q}, \mathcal{D});$ \\
        \For{$\mathbf{d} \in \mathbf{D}$}{
            $\mathcal{P}\cup Seg(\mathbf{d})$;\% passage segmentation; \\
        }
        \If {$\mathbf{q}$ is a training query}{
            \% cluster-based passage de-duplication for training query; \\
            $\mathcal{P} = \text{De-duplication}(\mathcal{P})$; \\
            \If {$CE(\cdot)$ is ready}{
                \% filtering out the query-passage pairs with high certainty;\\
                $\mathcal{P} = \text{Filter}(CE(\mathbf{q}, \mathcal{P}))$; \\
            }
        }
        \% passages of test queries are all annotated to alleviate the false negative issue;\\
        \For{$\mathbf{p}\in \mathcal{P}$}{
            \% annotation for query-passage pair;\\
            $\mathcal{L}\cup \mathcal{H}(\mathbf{q},\mathbf{p})$;\\
        }
    }
    return $\mathcal{L}$
}
\end{algorithm}

%% file: Sections/5-data-statistics.tex
\section{Data Statistics}
This section presents the data statistics of $\rm T^2Ranking$.

\begin{figure}
    \centering
    \includegraphics[width=\linewidth]{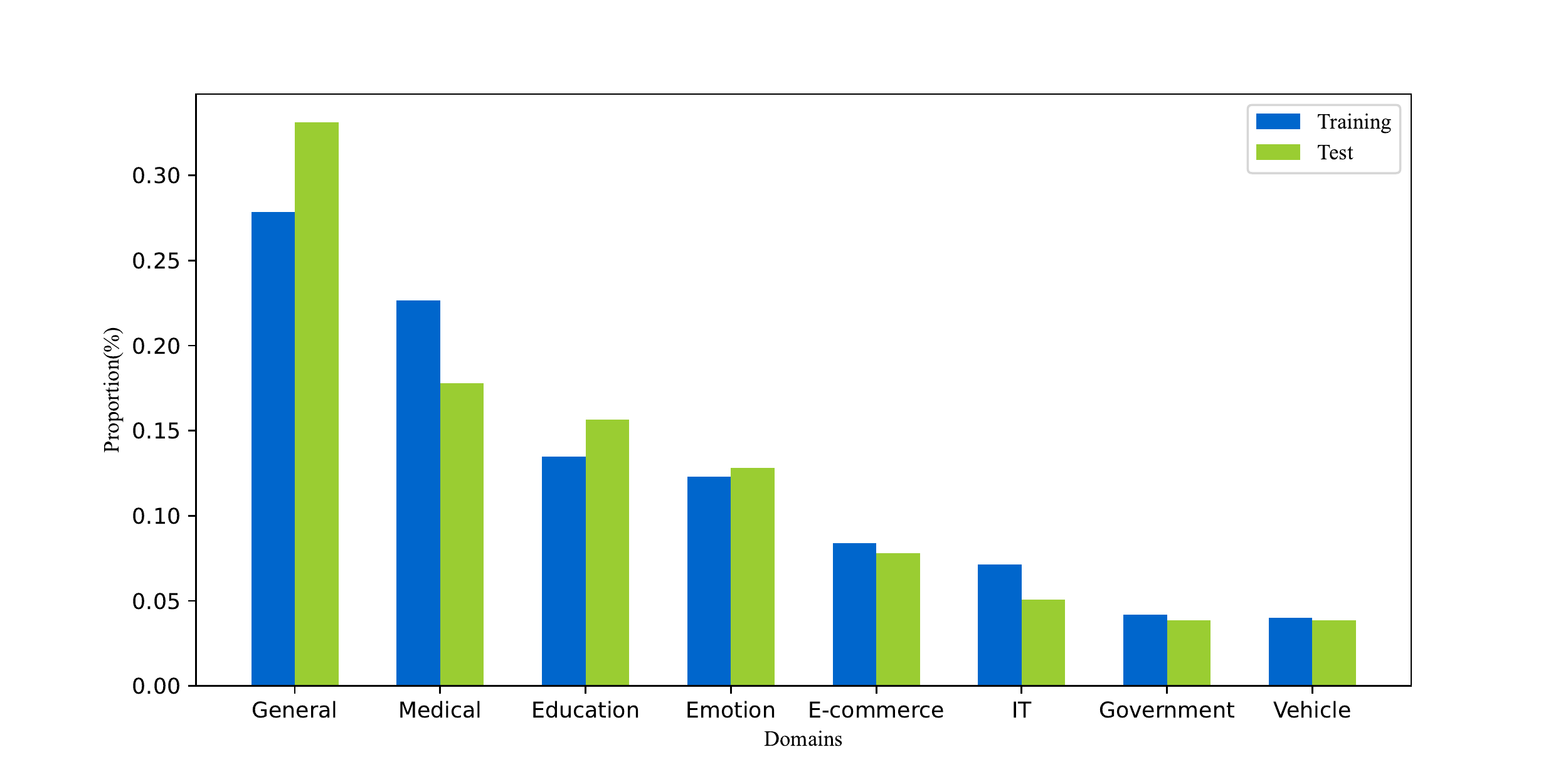}
    \caption{Domain statistics for the training and test queries in $\rm T^2Ranking$.}
    \label{fig:domain}
\end{figure}

\noindent \textbf{Query.} Table~\ref{tab:query} provides a summary of the statistics of queries in $\rm T^2Ranking$. The maximum and mean lengths of queries in the training and test sets are nearly identical. 
We further analyze the domain distribution of queries in the training and test sets, as demonstrated in Figure~\ref{fig:domain}. 
Domain tags are provided by the Sogou search engine.
The query domain distribution in the training and test sets is consistent, and the queries cover a broad range of domains.
We also demonstrate the diversity level of queries by resorting to the metric, intra-list similarity (ILS)~\cite{ziegler2005improving} which can be defined as
\begin{equation}
s(\textbf{q}_i, \textbf{q}_j)=\frac{\operatorname{BERT}(\textbf{q}_i)_{[c l s]} \cdot \operatorname{BERT}(\textbf{q}_j)_{[c l s]}}
{\| \operatorname{BERT}(\textbf{q}_i)_{[c l s]}\|\| \operatorname{BERT}(\textbf{q}_j)_{[cls]}\|},
\end{equation}
\begin{equation}
    \mathbf{I L S}_{\mathcal{Q}}=\frac{\sum_{i=1}^{|\mathcal{Q}|} \sum_{j=i+1}^{\mathcal{Q}} s(\textbf{q}_i, \textbf{q}_j)}{\sum_{i=1}^{\mathcal{Q}} \sum_{j=i+1}^{\mathcal{Q}} 1},
\end{equation}
where BERT~\cite{devlin2018bert} is a pre-trained language model that is often used as the backbone model for various tasks~\cite{nogueira2019multi,karpukhin2020dense,dong2021latent,dong2022incorporating, dong2022disentangled}.
A lower ILS score indicates a lower similarity between queries in the benchmark, thus indicating a higher level of diversity.
We calculated the ILS scores of $\rm T^2Ranking$, as well as those of several popular datasets, such as MSMARCO, Multi-CPR and $\rm DuReader_{retrieval}$. The results are shown in Table~\ref{tab:ils}. 
From the table, it is evident that the queries in $\rm T^2Ranking$ are more diverse, as indicated by a lower ILS score.
Note that, $\rm T^2Ranking$ compromises queries with higher diversity even than those in Multi-CPR, which contains queries from different vertical search applications.

\begin{table}[]
\caption{Statistic of queries in $\rm T^2Ranking$.}
\label{tab:query}
\begin{tabular}{l|c|c|c}
\hline
             & Quantity & Max. length & Mean.length \\ \hline
Training set & 258,042  & 40          & 11.1        \\ \hline
Test set  & 49,662   & 38          & 10.99       \\ \hline
\end{tabular}
\end{table}

\noindent \textbf{Document \& Passage.} $\rm T^2Ranking$ comprises passages extracted from 1,752,482 web documents, with a total of 2,303,643 passages after segmentation. On average, each web document is divided into 1.31 passages of which the mean length is 632.6.

\begin{figure}		
    \centering
    \subfloat[The distribution of relevance annotations in the training set.]{\includegraphics[width=0.45\linewidth]{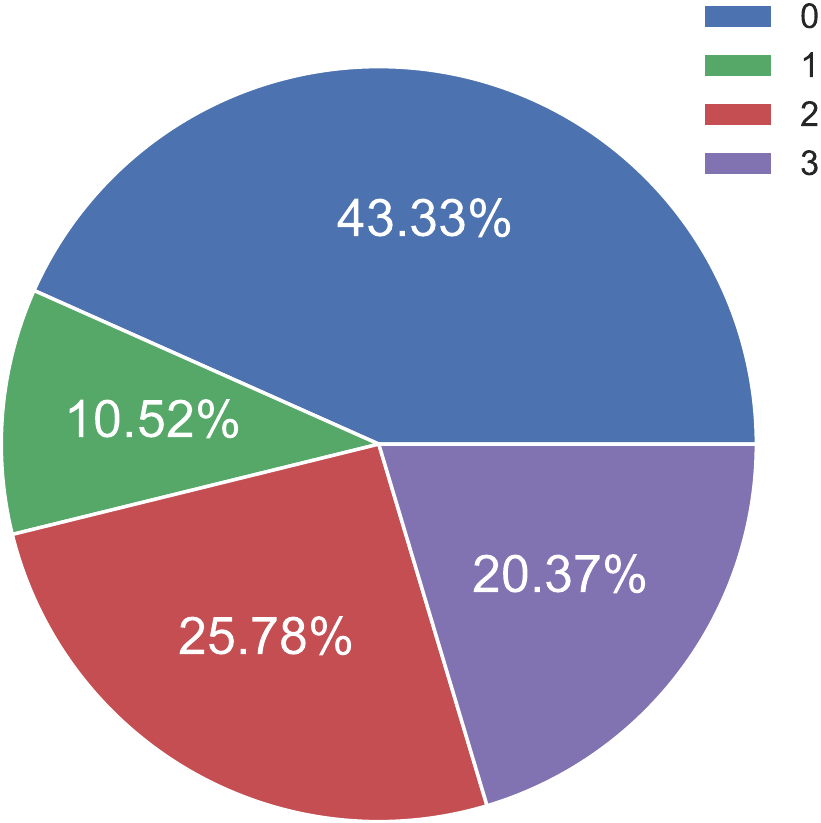}}\hspace{5mm}
    \subfloat[The distribution of relevance annotations in the test set.]{\includegraphics[width=0.45\linewidth]{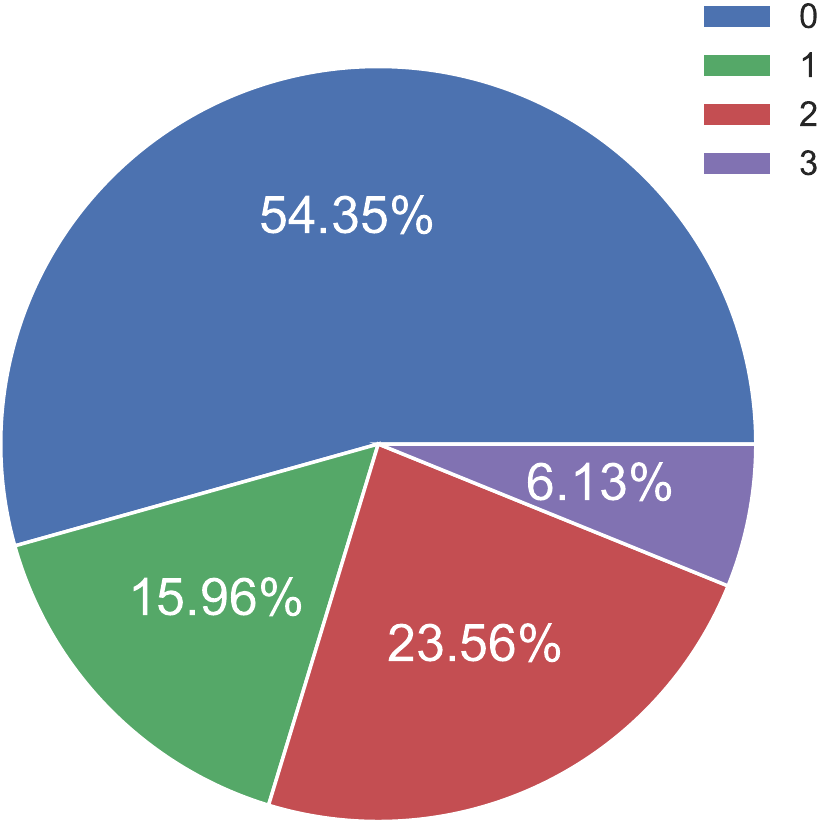}}
    \caption{Pie chart of the annotation distribution.}
    \label{fig:annotation}
\end{figure}

\noindent \textbf{Relevance Annotation.} 
We display the distribution of the 4-level relevance annotations in Figure~\ref{fig:annotation}. 
In the training set, on average, each query is annotated with 6.25 passages, while in the test set, each query is annotated with an average of 15.75 passages. 

\begin{table}[h]
\caption{ILS scores of different datasets. Lower ILS scores refer to higher diversity levels of queries.}
\label{tab:ils}
\begin{tabular}{c|c}
\toprule
  Dataset        & ILS Score  \\ \midrule
MS-MARCO~\cite{nguyen2016ms}  & 0.227 \\ 
Multi-CPR~\cite{long2022multi} & 0.186 \\ 
$\rm DuReader_{retrieval}$~\cite{qiu2022dureader_retrieval}  & 0.152 \\ \midrule
$\rm T^2Ranking$~(ours) & \textbf{0.144} \\ \bottomrule
\end{tabular}
\end{table}

%% file: Sections/6-experiment.tex

\section{Experiments and Results}
\label{section:experiment}
Consistent with modern information retrieval systems, the \textit{retrieval-then-re-ranking} paradigm is utilized in our experiments.
In this section, we examine the performance of commonly-used retrievers and re-rankers on $\rm T^2Ranking$. 

\begin{figure}
		\centering
		\includegraphics[width=\linewidth]{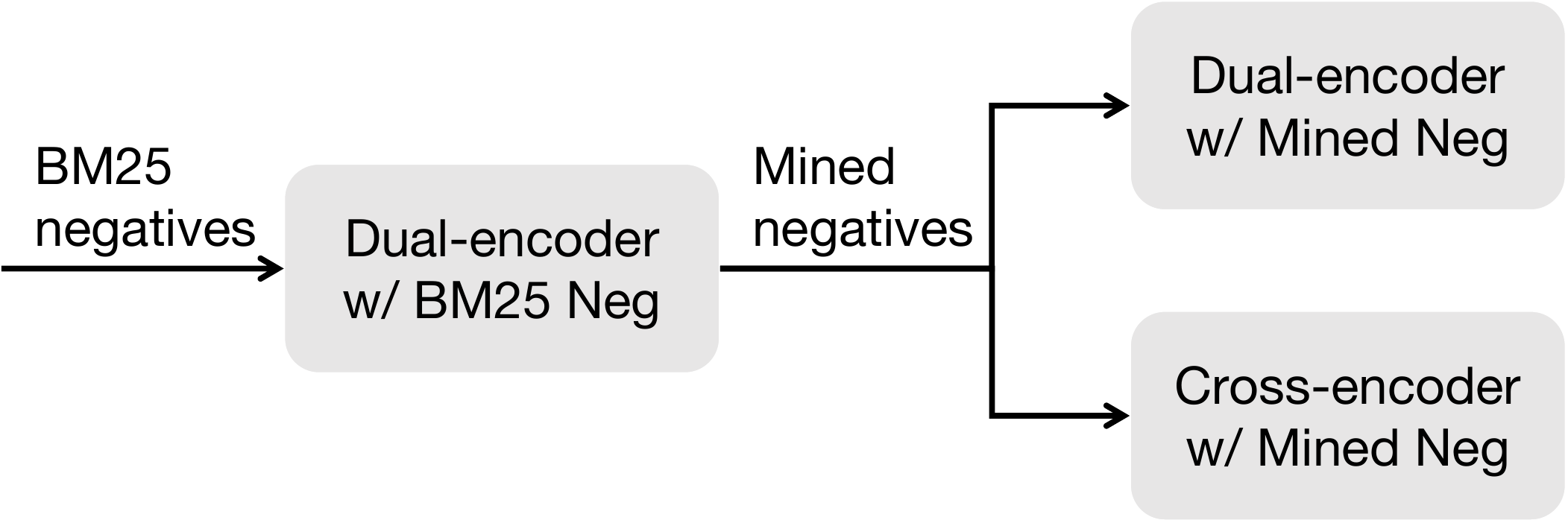}
		\caption{Illustration for the training process of baselines used in our experiments. First, we train a dual-encoder with BM25 negatives, which is similar to DPR~\cite{karpukhin2020dense}. Second, we train the dual-encoder and cross-encoder with the global negative sampling strategy proposed in several studies~\cite{long2022multi,qiu2022dureader_retrieval}.}
		\label{fig:trainingDenseRetrieval}
\end{figure}

\begin{figure}		
    \centering
    \subfloat[Dual-encoder.]{\includegraphics[width=0.45\linewidth]{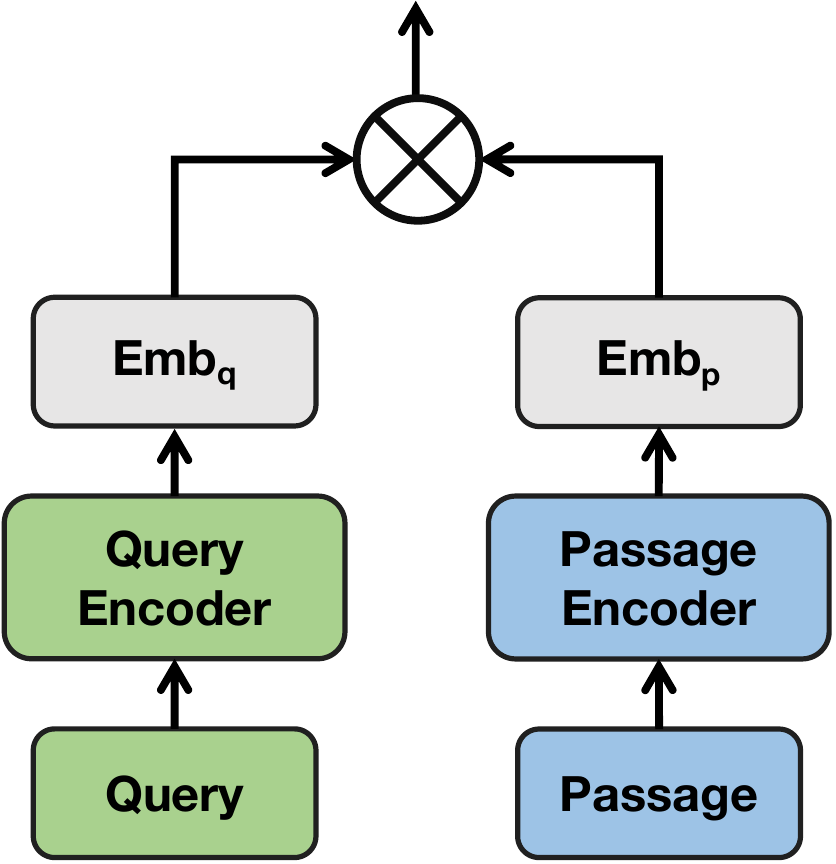}}\hspace{5mm}
    \subfloat[Cross-encoder.]{\includegraphics[width=0.45\linewidth]{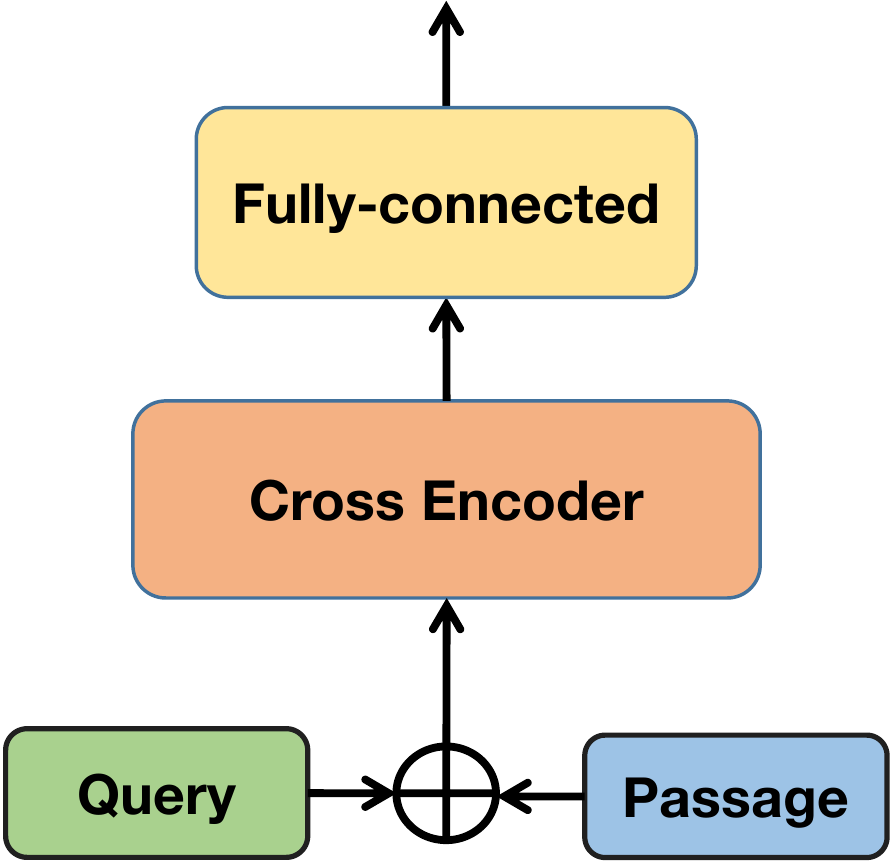}}
    \caption{Illustration for the architecture of dual-encoder and cross-encoder.}
    \label{fig:model}
\end{figure}

\subsection{Retrieval Performance}
\label{subsection:retrievalperformance}
\textbf{Baselines.} 
Existing retrieval models can be broadly divided into sparse retrieval models and dense retrieval models. 
Sparse retrieval models focus on exact matching signals to design a relevance scoring function, with BM25 being the most prominent and widely-utilized baseline due to its promising performance. 
Additionally, dense retrieval models leverage deep neural networks to learn low-dimensional dense embeddings for queries and documents. 
Generally, most existing dense retrieval methods adhere to the cascade training paradigm~\cite{qu2020rocketqa,qiu2022dureader_retrieval,long2022multi}. 
Therefore, to facilitate easier comparison in future studies on our dataset, we simplify the training process as illustrated in Figure~\ref{fig:trainingDenseRetrieval} as in~\cite{long2022multi,qiu2022dureader_retrieval}. 
Specifically, we utilize the dual-encoder (DE) as the architecture of dense retrieval models, which is illustrated in Figure~\ref{fig:model}(a).
The following methods are employed as our baselines to evaluate the retrieval performance on $\rm T^2Ranking$.
\begin{itemize}[nosep]
    \item \textbf{QL} (query likelihood)~\cite{ponte2017language} is a representative statistical language model that measures the relevance of passages by modeling the generation of a query.
    \item \textbf{BM25}~\cite{robertson2009probabilistic} is a widely-used sparse retrieval baseline.
    \item \textbf{DE w/ BM25 Neg} is equivalent to DPR~\cite{karpukhin2020dense}, which is the first work that uses the pre-trained language model as the backbone for the passage retrieval task.
    \item \textbf{DE w/ Mined Neg} enhance the performance of DPR by sampling hard negatives globally from the entire corpus as in ANCE~\cite{xiong2020approximate} and RocketQA~\cite{qu2020rocketqa}.
    \item \textbf{DPTDR}~\cite{tang2022dptdr} is the first work that employs prompt tuning for dense retrieval.
\end{itemize}
Among them, QL and BM25 are sparse retrieval models, whereas the others are dense retrieval models

\noindent \textbf{Implementation details.} BM25 is implemented by Pyserini~\cite{lin2021pyserini} with default parameters. The dual-encoder models are implemented by the deep learning framework PyTorch on up to 8 NVIDIA Tesla A100 GPUs (with 80G RAM). We use the off-the-shelf Chinese BERT$_{\text{base}}$ to initialize the dual-encoder. The maximal length of queries and passages are set to 32 and 256, respectively. The negatives are sampled from the top 200 passages recalled by BM25 or DE w/ BM25 Neg. The ratio of positive:negative is set to 1:1. We train the dual-encoder for 100 epochs with a learning rate of 3e-5.

\noindent \textbf{Metrics.} The following evaluation metrics are used in our experiments to examine the retrieval performance of baselines on $\rm T^2Ranking$: (1) Mean Reciprocal Rank for the top 10 retrieved passages (MRR@10), (2) Recall for the top-$K$ retrieved passages (Recall@$K$). Notably, for the retrieval task, we consider \textbf{Level-2} and \textbf{Level-3} passages as relevant passages, and all other passages are regarded as irrelevant passages. For a comprehensive comparison, we report Recall@50 and Recall@1K on the test queries. 
Following the evaluation settings of MS-MARCO and $\rm DuReader_{retrieval}$, MRR is defined as the average of the reciprocal ranks of the \textit{first} relevant passage for a set of queries. 
The MRR is a value between 0 and 1, with a higher value indicating that the system is better at ranking the most relevant passage higher in the list. 
Meanwhile, Recall is defined as the fraction of relevant passages that are retrieved among all relevant passages, also with a value between 0 and 1, where a higher value indicates that the system is better at retrieving all relevant passages. 
MRR and Recall measure different aspects of retrieval performance.
MRR@$K$ and Recall@$K$ can be depicted as:
\begin{equation}
    \label{eq:mrr}
    MRR@K = \frac{1}{|\mathcal{Q}|}\sum_{q\in \mathcal{Q}}\frac{\mathbf{I}(rank\leq K)}{rank},
\end{equation}
\begin{equation}
    \label{eq:recall}
    Recall@K = \frac{\mathbf{I}(rank_{p}^{\mathcal{K}^q}\leq K)}{\sum_{q\in \mathcal{Q}}\sum_{p\in R_{q}} 1}.
\end{equation}
where $\mathbf{I(\cdot)}$ is a indicator function. The $rank$ in Eq.~\ref{eq:mrr} denotes the position of the \textit{first} relevant passage in the retrieved candidates of query $q$. The $R_q$ and $rank_{p}^{\mathcal{K}^q}$ represent the relevant passages of query $q$ and the position of passage $p$ in the candidate list $\mathcal{K}^q$.

\begin{table}[]
    \centering
    \caption{Performance of retrieval models on the test set of $\rm T^2Ranking$.}
    \begin{tabular}{lcccc}
        \toprule & MRR@10 & Recall @50 & Recall@1K \\
        \midrule 
        QL & $.2803$ & $.3915$ & $.6858$ \\
        BM25 & $.3579$  & $.4918$ & $.7426$ \\
         DE w/ BM25 Neg & $.4877$  & $.7123$ & $.9104$ \\
         DE w/ Mined Neg & $.5191$  & $.7357$ & $.9147$ \\
         DPTDR & $\textbf{.5285}$ & $\textbf{.7423}$ & $\textbf{.9211}$ \\
        \bottomrule
    \end{tabular}
    \label{tab:retrievalPerformance}
\end{table}

\noindent \textbf{Retrieval performance.} 
We report the retrieval performance of baselines in Table~\ref{tab:retrievalPerformance}. 
Compared to the traditional sparse retrieval method BM25, dual-encoder models significantly boost the retrieval performance on our dataset. 
The improvement can be attributed to the integration of two distinct sources of knowledge, i.e., latent knowledge obtained through unsupervised pre-training of language models on a massive corpus and relevance knowledge acquired through supervised training on our large-scale annotated dataset.
Equipped with the strategy of negative mining proposed in recent studies~\cite{xiong2020approximate}, the retrieval performance of dual-encoder models could be further improved on $\rm T^2Ranking$. 
It is worth noting that the Recall@$K$ metrics observed in $\rm T^2Ranking$ are lower than those reported in other benchmarks with coarse-grained annotations. 
For instance, the Recall@50 of BM25 is .601 and .700 on MS-MARCO-DEV Passage and $\rm DuReader_{retrieval}$, respectively, and 0.4918 on our dataset.
In the test set of $\rm T^2Ranking$, we have a greater number of passages annotated with fine-grained relevance labels, leading to a 4.74 average positive paragraph per query, which makes the retrieval task more difficult and eases the false negative problem to some extent.
This highlights the challenging nature of $\rm T^2Ranking$ and the potential for further improvement in the future.

\subsection{Re-ranking Performance}
\noindent \textbf{Baselines.} 
Due to the smaller number of passages considered by re-rankers, they tend to use the cross-encoder architecture rather than the dual-encoder architecture. 
The cross-encoder approach allows for a more detailed interaction between queries and documents, resulting in better performance, although at the expense of lower efficiency.
We report the re-ranking performance of the cross-encoder model, which is trained on the hard negatives mined from the entire corpus, as depicted in Figure~\ref{fig:trainingDenseRetrieval}. The architecture of cross-encoder is illustrated in Figure~\ref{fig:model}(b).

\noindent \textbf{Implementation details.} 
The cross-encoder is implemented in the same experimental environment as the dual-encoder, with a maximum input length of 288. Negatives are sampled from the top 256 passages retrieved by the dual-encoder, and a positive-to-negative ratio of 1:128 is set. The cross-encoder is then trained for 5 epochs with a learning rate of 3e-5.

\noindent \textbf{Metrics.} To evaluate the re-ranking performance of the cross-encoder, we use two ranking metrics: MRR@10 and nDCG@$K$. 
In the test set of $\rm T^2Ranking$, the average number of annotated passages per query is 15.7, with a maximum of 100 annotated passages. 
We report nDCG@20 and nDCG@100 on the test queries. 
nDCG@$K$ normalizes DCG@$K$ by dividing DCG@$K$ by the iDCG@$K$, which is the DCG@$K$ of ideal ordering of the passages. 
DCG@$K$ discounts the graded relevance value of a passage according to the rank that it appears at, which can be defined as:
\begin{equation}
\label{eq:dcg}
DCG@K=\frac{1}{|\mathcal{Q}|}\sum_{q\in \mathcal{Q}} \sum_{p\in \mathcal{K}^q} \frac{L\left(p\right)}{\log_2(rank_p^{\mathcal{K}^q}+1)},
\end{equation}
\begin{equation}
    \label{eq:ndcg}
    nDCG@K=\frac{DCG@K}{iDCG@K},
\end{equation}
where $L(p)$ is the graded relevance of passage $p$.

\noindent \textbf{Re-ranking performance.} 
The re-ranking performance of the cross-encoder is shown in Table~\ref{tab:rerankingPerformance}. The results indicate that re-ranking the candidates retrieved by the dual-encoder significantly outperforms re-ranking the candidates retrieved using the BM25 method. 
The improved performance is attributed to the higher recall rate achieved by the dual-encoder method, which is consistent with previous studies conducted on other benchmarks~\cite{qiu2022dureader_retrieval,long2022multi}.
The re-ranking performance on $\rm T^2Ranking$, however, is lower compared to other benchmarks~\cite{qiu2022dureader_retrieval,long2022multi}.
This can be explained by the presence of more fine-grained annotated relevant passages and queries with higher diversities in $\rm T^2Ranking$, which makes it a more challenging benchmark but also provides a more accurate reflection of re-ranking performance.

\label{subsection:rankingperformance}
\begin{table}[]
    \centering
    \caption{Performance of cross-encoder with mined negatives on the test set of $\rm T^2Ranking$.}
    \begin{tabular}{lccc}
    \toprule Candidates & MRR@10 & nDCG@20 & nDCG@100 \\
    \midrule  
     BM25's top-1000 psg. & $.5184$ & $.4401$ & $.4696$ \\
     DE's top-1000 psg. & $.5520$ & $.5149$ & $.5571$ \\
    \bottomrule
    \end{tabular}
    \label{tab:rerankingPerformance}
\end{table}


%% file: Sections/7-conclusion.tex

\section{Conclusion}
\label{section:conclusion}
In this study, we introduce $\rm T^2Ranking$, a large-scale benchmark for Chinese passage ranking that involves both retrieval and re-ranking tasks.
To construct a high-quality dataset, we leverage various strategies, including model-based passage segmentation, clustering-based passage de-duplication and active learning-based data sampling.
Specifically, we adopt a model-based method for passage segmentation in $\rm T^2Ranking$, which aims to maximize the preservation of complete semantics in each passage.
To balance the efficiency of annotation with the diversity of annotated query-passage pairs, we incorporate a clustering-based technique in $\rm T^2Ranking$ to remove highly similar passages retrieved by a specific query, which helps streamline the annotation process without compromising the overall quality of the dataset.
The adoption of an active learning strategy in the construction of $\rm T^2Ranking$ enhances the efficiency of annotating more informative training samples. The active learning framework enables the dataset to be continuously updated with the most valuable samples while minimizing the number of annotations required to achieve optimal performance.
Furthermore, to ensure high-quality annotation, expert annotators are involved in the implementation of a 4-level fine-grained annotation scheme for both the training and test sets in $\rm T^2Ranking$. 
This scheme allows for more nuanced modeling of IR models during training and a more precise evaluation of the models during testing.
In summary, $\rm T^2Ranking$ encompasses over 300K queries and more than 2M unique passages, with around 2.4 million query-passage pairs annotated with fine-grained relevance labels by expert annotators.
To the best of our knowledge, $\rm T^2Ranking$ is the largest Chinese benchmark with fine-grained annotation for passage ranking. 
We believe that this dataset will make a significant contribution to the IR community and the advancement of IR technology.

%% file: SIGIR2023.bbl

\begin{thebibliography}{31}


\ifx \showCODEN    \undefined \def \showCODEN     #1{\unskip}     \fi
\ifx \showDOI      \undefined \def \showDOI       #1{#1}\fi
\ifx \showISBNx    \undefined \def \showISBNx     #1{\unskip}     \fi
\ifx \showISBNxiii \undefined \def \showISBNxiii  #1{\unskip}     \fi
\ifx \showISSN     \undefined \def \showISSN      #1{\unskip}     \fi
\ifx \showLCCN     \undefined \def \showLCCN      #1{\unskip}     \fi
\ifx \shownote     \undefined \def \shownote      #1{#1}          \fi
\ifx \showarticletitle \undefined \def \showarticletitle #1{#1}   \fi
\ifx \showURL      \undefined \def \showURL       {\relax}        \fi
\providecommand\bibfield[2]{#2}
\providecommand\bibinfo[2]{#2}
\providecommand\natexlab[1]{#1}
\providecommand\showeprint[2][]{arXiv:#2}

\bibitem[Aktolga et~al\mbox{.}(2011)]%
        {aktolga2011passage}
\bibfield{author}{\bibinfo{person}{Elif Aktolga}, \bibinfo{person}{James
  Allan}, {and} \bibinfo{person}{David~A Smith}.}
  \bibinfo{year}{2011}\natexlab{}.
\newblock \showarticletitle{Passage reranking for question answering using
  syntactic structures and answer types}. In \bibinfo{booktitle}{\emph{Advances
  in Information Retrieval: 33rd European Conference on IR Research, ECIR 2011,
  Dublin, Ireland, April 18-21, 2011. Proceedings 33}}. Springer,
  \bibinfo{pages}{617--628}.
\newblock


\bibitem[Arabzadeh et~al\mbox{.}(2022)]%
        {arabzadeh2022shallow}
\bibfield{author}{\bibinfo{person}{Negar Arabzadeh}, \bibinfo{person}{Alexandra
  Vtyurina}, \bibinfo{person}{Xinyi Yan}, {and} \bibinfo{person}{Charles~LA
  Clarke}.} \bibinfo{year}{2022}\natexlab{}.
\newblock \showarticletitle{Shallow pooling for sparse labels}.
\newblock \bibinfo{journal}{\emph{Information Retrieval Journal}}
  \bibinfo{volume}{25}, \bibinfo{number}{4} (\bibinfo{year}{2022}),
  \bibinfo{pages}{365--385}.
\newblock


\bibitem[Bonifacio et~al\mbox{.}(2021)]%
        {bonifacio2021mmarco}
\bibfield{author}{\bibinfo{person}{Luiz Bonifacio}, \bibinfo{person}{Vitor
  Jeronymo}, \bibinfo{person}{Hugo~Queiroz Abonizio}, \bibinfo{person}{Israel
  Campiotti}, \bibinfo{person}{Marzieh Fadaee}, \bibinfo{person}{Roberto
  Lotufo}, {and} \bibinfo{person}{Rodrigo Nogueira}.}
  \bibinfo{year}{2021}\natexlab{}.
\newblock \showarticletitle{mmarco: A multilingual version of the ms marco
  passage ranking dataset}.
\newblock \bibinfo{journal}{\emph{arXiv preprint arXiv:2108.13897}}
  (\bibinfo{year}{2021}).
\newblock


\bibitem[Cheng et~al\mbox{.}(2023)]%
        {cheng2023layout}
\bibfield{author}{\bibinfo{person}{Anfeng Cheng}, \bibinfo{person}{Yiding Liu},
  \bibinfo{person}{Weibin Li}, \bibinfo{person}{Qian Dong},
  \bibinfo{person}{Shuaiqiang Wang}, \bibinfo{person}{Zhengjie Huang},
  \bibinfo{person}{Shikun Feng}, \bibinfo{person}{Zhicong Cheng}, {and}
  \bibinfo{person}{Dawei Yin}.} \bibinfo{year}{2023}\natexlab{}.
\newblock \showarticletitle{Layout-aware Webpage Quality Assessment}.
\newblock \bibinfo{journal}{\emph{arXiv preprint arXiv:2301.12152}}
  (\bibinfo{year}{2023}).
\newblock


\bibitem[Craswell et~al\mbox{.}(2021)]%
        {craswell2021trec}
\bibfield{author}{\bibinfo{person}{Nick Craswell}, \bibinfo{person}{Bhaskar
  Mitra}, \bibinfo{person}{Emine Yilmaz}, \bibinfo{person}{Daniel Campos},
  \bibinfo{person}{Ellen~M Voorhees}, {and} \bibinfo{person}{Ian Soboroff}.}
  \bibinfo{year}{2021}\natexlab{}.
\newblock \showarticletitle{TREC deep learning track: Reusable test collections
  in the large data regime}. In \bibinfo{booktitle}{\emph{Proceedings of the
  44th international ACM SIGIR conference on research and development in
  information retrieval}}. \bibinfo{pages}{2369--2375}.
\newblock


\bibitem[Dietz et~al\mbox{.}(2017)]%
        {dietz2017trec}
\bibfield{author}{\bibinfo{person}{Laura Dietz}, \bibinfo{person}{Manisha
  Verma}, \bibinfo{person}{Filip Radlinski}, {and} \bibinfo{person}{Nick
  Craswell}.} \bibinfo{year}{2017}\natexlab{}.
\newblock \showarticletitle{TREC Complex Answer Retrieval Overview.}. In
  \bibinfo{booktitle}{\emph{TREC}}.
\newblock


\bibitem[Dong et~al\mbox{.}(2022a)]%
        {dong2022incorporating}
\bibfield{author}{\bibinfo{person}{Qian Dong}, \bibinfo{person}{Yiding Liu},
  \bibinfo{person}{Suqi Cheng}, \bibinfo{person}{Shuaiqiang Wang},
  \bibinfo{person}{Zhicong Cheng}, \bibinfo{person}{Shuzi Niu}, {and}
  \bibinfo{person}{Dawei Yin}.} \bibinfo{year}{2022}\natexlab{a}.
\newblock \showarticletitle{Incorporating Explicit Knowledge in Pre-trained
  Language Models for Passage Re-ranking}.
\newblock \bibinfo{journal}{\emph{arXiv preprint arXiv:2204.11673}}
  (\bibinfo{year}{2022}).
\newblock


\bibitem[Dong and Niu(2021)]%
        {dong2021latent}
\bibfield{author}{\bibinfo{person}{Qian Dong} {and} \bibinfo{person}{Shuzi
  Niu}.} \bibinfo{year}{2021}\natexlab{}.
\newblock \showarticletitle{Latent Graph Recurrent Network for Document
  Ranking}. In \bibinfo{booktitle}{\emph{Database Systems for Advanced
  Applications: 26th International Conference, DASFAA 2021, Taipei, Taiwan,
  April 11--14, 2021, Proceedings, Part II 26}}. Springer,
  \bibinfo{pages}{88--103}.
\newblock


\bibitem[Dong et~al\mbox{.}(2022b)]%
        {dong2022disentangled}
\bibfield{author}{\bibinfo{person}{Qian Dong}, \bibinfo{person}{Shuzi Niu},
  \bibinfo{person}{Tao Yuan}, {and} \bibinfo{person}{Yucheng Li}.}
  \bibinfo{year}{2022}\natexlab{b}.
\newblock \showarticletitle{Disentangled graph recurrent network for document
  ranking}.
\newblock \bibinfo{journal}{\emph{Data Science and Engineering}}
  \bibinfo{volume}{7}, \bibinfo{number}{1} (\bibinfo{year}{2022}),
  \bibinfo{pages}{30--43}.
\newblock


\bibitem[Fan et~al\mbox{.}(2022)]%
        {fan2022pre}
\bibfield{author}{\bibinfo{person}{Yixing Fan}, \bibinfo{person}{Xiaohui Xie},
  \bibinfo{person}{Yinqiong Cai}, \bibinfo{person}{Jia Chen},
  \bibinfo{person}{Xinyu Ma}, \bibinfo{person}{Xiangsheng Li},
  \bibinfo{person}{Ruqing Zhang}, \bibinfo{person}{Jiafeng Guo},
  {et~al\mbox{.}}} \bibinfo{year}{2022}\natexlab{}.
\newblock \showarticletitle{Pre-training methods in information retrieval}.
\newblock \bibinfo{journal}{\emph{Foundations and Trends{\textregistered} in
  Information Retrieval}} \bibinfo{volume}{16}, \bibinfo{number}{3}
  (\bibinfo{year}{2022}), \bibinfo{pages}{178--317}.
\newblock


\bibitem[Joshi et~al\mbox{.}(2017)]%
        {joshi2017triviaqa}
\bibfield{author}{\bibinfo{person}{Mandar Joshi}, \bibinfo{person}{Eunsol
  Choi}, \bibinfo{person}{Daniel~S Weld}, {and} \bibinfo{person}{Luke
  Zettlemoyer}.} \bibinfo{year}{2017}\natexlab{}.
\newblock \showarticletitle{Triviaqa: A large scale distantly supervised
  challenge dataset for reading comprehension}.
\newblock \bibinfo{journal}{\emph{arXiv preprint arXiv:1705.03551}}
  (\bibinfo{year}{2017}).
\newblock


\bibitem[Karpukhin et~al\mbox{.}(2020)]%
        {karpukhin2020dense}
\bibfield{author}{\bibinfo{person}{Vladimir Karpukhin}, \bibinfo{person}{Barlas
  O{\u{g}}uz}, \bibinfo{person}{Sewon Min}, \bibinfo{person}{Patrick Lewis},
  \bibinfo{person}{Ledell Wu}, \bibinfo{person}{Sergey Edunov},
  \bibinfo{person}{Danqi Chen}, {and} \bibinfo{person}{Wen-tau Yih}.}
  \bibinfo{year}{2020}\natexlab{}.
\newblock \showarticletitle{Dense passage retrieval for open-domain question
  answering}.
\newblock \bibinfo{journal}{\emph{arXiv preprint arXiv:2004.04906}}
  (\bibinfo{year}{2020}).
\newblock


\bibitem[Kenton and Toutanova(2019)]%
        {devlin2018bert}
\bibfield{author}{\bibinfo{person}{Jacob Devlin Ming-Wei~Chang Kenton} {and}
  \bibinfo{person}{Lee~Kristina Toutanova}.} \bibinfo{year}{2019}\natexlab{}.
\newblock \showarticletitle{Bert: Pre-training of deep bidirectional
  transformers for language understanding}. In
  \bibinfo{booktitle}{\emph{Proceedings of naacL-HLT}},
  Vol.~\bibinfo{volume}{1}. \bibinfo{pages}{2}.
\newblock


\bibitem[Lin et~al\mbox{.}(2021)]%
        {lin2021pyserini}
\bibfield{author}{\bibinfo{person}{Jimmy Lin}, \bibinfo{person}{Xueguang Ma},
  \bibinfo{person}{Sheng-Chieh Lin}, \bibinfo{person}{Jheng-Hong Yang},
  \bibinfo{person}{Ronak Pradeep}, {and} \bibinfo{person}{Rodrigo Nogueira}.}
  \bibinfo{year}{2021}\natexlab{}.
\newblock \showarticletitle{Pyserini: A Python toolkit for reproducible
  information retrieval research with sparse and dense representations}. In
  \bibinfo{booktitle}{\emph{Proceedings of the 44th International ACM SIGIR
  Conference on Research and Development in Information Retrieval}}.
  \bibinfo{pages}{2356--2362}.
\newblock


\bibitem[Long et~al\mbox{.}(2022)]%
        {long2022multi}
\bibfield{author}{\bibinfo{person}{Dingkun Long}, \bibinfo{person}{Qiong Gao},
  \bibinfo{person}{Kuan Zou}, \bibinfo{person}{Guangwei Xu},
  \bibinfo{person}{Pengjun Xie}, \bibinfo{person}{Ruijie Guo},
  \bibinfo{person}{Jian Xu}, \bibinfo{person}{Guanjun Jiang},
  \bibinfo{person}{Luxi Xing}, {and} \bibinfo{person}{Ping Yang}.}
  \bibinfo{year}{2022}\natexlab{}.
\newblock \showarticletitle{Multi-CPR: A Multi Domain Chinese Dataset for
  Passage Retrieval}. In \bibinfo{booktitle}{\emph{Proceedings of the 45th
  International ACM SIGIR Conference on Research and Development in Information
  Retrieval}}. \bibinfo{pages}{3046--3056}.
\newblock


\bibitem[Nguyen et~al\mbox{.}(2016)]%
        {nguyen2016ms}
\bibfield{author}{\bibinfo{person}{Tri Nguyen}, \bibinfo{person}{Mir
  Rosenberg}, \bibinfo{person}{Xia Song}, \bibinfo{person}{Jianfeng Gao},
  \bibinfo{person}{Saurabh Tiwary}, \bibinfo{person}{Rangan Majumder}, {and}
  \bibinfo{person}{Li Deng}.} \bibinfo{year}{2016}\natexlab{}.
\newblock \showarticletitle{MS MARCO: A human generated machine reading
  comprehension dataset}. In \bibinfo{booktitle}{\emph{CoCo@ NIPs}}.
\newblock


\bibitem[Nishida et~al\mbox{.}(2018)]%
        {nishida2018retrieve}
\bibfield{author}{\bibinfo{person}{Kyosuke Nishida}, \bibinfo{person}{Itsumi
  Saito}, \bibinfo{person}{Atsushi Otsuka}, \bibinfo{person}{Hisako Asano},
  {and} \bibinfo{person}{Junji Tomita}.} \bibinfo{year}{2018}\natexlab{}.
\newblock \showarticletitle{Retrieve-and-read: Multi-task learning of
  information retrieval and reading comprehension}. In
  \bibinfo{booktitle}{\emph{Proceedings of the 27th ACM international
  conference on information and knowledge management}}.
  \bibinfo{pages}{647--656}.
\newblock


\bibitem[Nogueira et~al\mbox{.}(2019)]%
        {nogueira2019multi}
\bibfield{author}{\bibinfo{person}{Rodrigo Nogueira}, \bibinfo{person}{Wei
  Yang}, \bibinfo{person}{Kyunghyun Cho}, {and} \bibinfo{person}{Jimmy Lin}.}
  \bibinfo{year}{2019}\natexlab{}.
\newblock \showarticletitle{Multi-stage document ranking with BERT}.
\newblock \bibinfo{journal}{\emph{arXiv preprint arXiv:1910.14424}}
  (\bibinfo{year}{2019}).
\newblock


\bibitem[Ponte and Croft(2017)]%
        {ponte2017language}
\bibfield{author}{\bibinfo{person}{Jay~M Ponte} {and} \bibinfo{person}{W~Bruce
  Croft}.} \bibinfo{year}{2017}\natexlab{}.
\newblock \showarticletitle{A language modeling approach to information
  retrieval}. In \bibinfo{booktitle}{\emph{ACM SIGIR Forum}},
  Vol.~\bibinfo{volume}{51}. ACM New York, NY, USA, \bibinfo{pages}{202--208}.
\newblock


\bibitem[Qiu et~al\mbox{.}(2022)]%
        {qiu2022dureader_retrieval}
\bibfield{author}{\bibinfo{person}{Yifu Qiu}, \bibinfo{person}{Hongyu Li},
  \bibinfo{person}{Yingqi Qu}, \bibinfo{person}{Ying Chen},
  \bibinfo{person}{Qiaoqiao She}, \bibinfo{person}{Jing Liu},
  \bibinfo{person}{Hua Wu}, {and} \bibinfo{person}{Haifeng Wang}.}
  \bibinfo{year}{2022}\natexlab{}.
\newblock \showarticletitle{DuReader\_retrieval: A Large-scale Chinese
  Benchmark for Passage Retrieval from Web Search Engine}.
\newblock \bibinfo{journal}{\emph{arXiv preprint arXiv:2203.10232}}
  (\bibinfo{year}{2022}).
\newblock


\bibitem[Qu et~al\mbox{.}(2020)]%
        {qu2020rocketqa}
\bibfield{author}{\bibinfo{person}{Yingqi Qu}, \bibinfo{person}{Yuchen Ding},
  \bibinfo{person}{Jing Liu}, \bibinfo{person}{Kai Liu},
  \bibinfo{person}{Ruiyang Ren}, \bibinfo{person}{Wayne~Xin Zhao},
  \bibinfo{person}{Daxiang Dong}, \bibinfo{person}{Hua Wu}, {and}
  \bibinfo{person}{Haifeng Wang}.} \bibinfo{year}{2020}\natexlab{}.
\newblock \showarticletitle{RocketQA: An optimized training approach to dense
  passage retrieval for open-domain question answering}.
\newblock \bibinfo{journal}{\emph{arXiv preprint arXiv:2010.08191}}
  (\bibinfo{year}{2020}).
\newblock


\bibitem[Ren et~al\mbox{.}(2021)]%
        {ren2021survey}
\bibfield{author}{\bibinfo{person}{Pengzhen Ren}, \bibinfo{person}{Yun Xiao},
  \bibinfo{person}{Xiaojun Chang}, \bibinfo{person}{Po-Yao Huang},
  \bibinfo{person}{Zhihui Li}, \bibinfo{person}{Brij~B Gupta},
  \bibinfo{person}{Xiaojiang Chen}, {and} \bibinfo{person}{Xin Wang}.}
  \bibinfo{year}{2021}\natexlab{}.
\newblock \showarticletitle{A survey of deep active learning}.
\newblock \bibinfo{journal}{\emph{ACM computing surveys (CSUR)}}
  \bibinfo{volume}{54}, \bibinfo{number}{9} (\bibinfo{year}{2021}),
  \bibinfo{pages}{1--40}.
\newblock


\bibitem[Robertson et~al\mbox{.}(2009)]%
        {robertson2009probabilistic}
\bibfield{author}{\bibinfo{person}{Stephen Robertson}, \bibinfo{person}{Hugo
  Zaragoza}, {et~al\mbox{.}}} \bibinfo{year}{2009}\natexlab{}.
\newblock \showarticletitle{The probabilistic relevance framework: BM25 and
  beyond}.
\newblock \bibinfo{journal}{\emph{Foundations and Trends{\textregistered} in
  Information Retrieval}} \bibinfo{volume}{3}, \bibinfo{number}{4}
  (\bibinfo{year}{2009}), \bibinfo{pages}{333--389}.
\newblock


\bibitem[Roitero et~al\mbox{.}(2018)]%
        {roitero2018fine}
\bibfield{author}{\bibinfo{person}{Kevin Roitero}, \bibinfo{person}{Eddy
  Maddalena}, \bibinfo{person}{Gianluca Demartini}, {and}
  \bibinfo{person}{Stefano Mizzaro}.} \bibinfo{year}{2018}\natexlab{}.
\newblock \showarticletitle{On fine-grained relevance scales}. In
  \bibinfo{booktitle}{\emph{The 41st International ACM SIGIR Conference on
  Research \& Development in Information Retrieval}}.
  \bibinfo{pages}{675--684}.
\newblock


\bibitem[Tang et~al\mbox{.}(2022)]%
        {tang2022dptdr}
\bibfield{author}{\bibinfo{person}{Zhengyang Tang}, \bibinfo{person}{Benyou
  Wang}, {and} \bibinfo{person}{Ting Yao}.} \bibinfo{year}{2022}\natexlab{}.
\newblock \showarticletitle{DPTDR: Deep Prompt Tuning for Dense Passage
  Retrieval}.
\newblock \bibinfo{journal}{\emph{arXiv preprint arXiv:2208.11503}}
  (\bibinfo{year}{2022}).
\newblock


\bibitem[Ward~Jr(1963)]%
        {ward1963hierarchical}
\bibfield{author}{\bibinfo{person}{Joe~H Ward~Jr}.}
  \bibinfo{year}{1963}\natexlab{}.
\newblock \showarticletitle{Hierarchical grouping to optimize an objective
  function}.
\newblock \bibinfo{journal}{\emph{Journal of the American statistical
  association}} \bibinfo{volume}{58}, \bibinfo{number}{301}
  (\bibinfo{year}{1963}), \bibinfo{pages}{236--244}.
\newblock


\bibitem[Wu et~al\mbox{.}(2020)]%
        {wu2020leveraging}
\bibfield{author}{\bibinfo{person}{Zhijing Wu}, \bibinfo{person}{Jiaxin Mao},
  \bibinfo{person}{Yiqun Liu}, \bibinfo{person}{Jingtao Zhan},
  \bibinfo{person}{Yukun Zheng}, \bibinfo{person}{Min Zhang}, {and}
  \bibinfo{person}{Shaoping Ma}.} \bibinfo{year}{2020}\natexlab{}.
\newblock \showarticletitle{Leveraging passage-level cumulative gain for
  document ranking}. In \bibinfo{booktitle}{\emph{Proceedings of The Web
  Conference 2020}}. \bibinfo{pages}{2421--2431}.
\newblock


\bibitem[Xiong et~al\mbox{.}(2020)]%
        {xiong2020approximate}
\bibfield{author}{\bibinfo{person}{Lee Xiong}, \bibinfo{person}{Chenyan Xiong},
  \bibinfo{person}{Ye Li}, \bibinfo{person}{Kwok-Fung Tang},
  \bibinfo{person}{Jialin Liu}, \bibinfo{person}{Paul Bennett},
  \bibinfo{person}{Junaid Ahmed}, {and} \bibinfo{person}{Arnold Overwijk}.}
  \bibinfo{year}{2020}\natexlab{}.
\newblock \showarticletitle{Approximate nearest neighbor negative contrastive
  learning for dense text retrieval}.
\newblock \bibinfo{journal}{\emph{arXiv preprint arXiv:2007.00808}}
  (\bibinfo{year}{2020}).
\newblock


\bibitem[Zhang et~al\mbox{.}(2018)]%
        {zhang2018relevance}
\bibfield{author}{\bibinfo{person}{Junqi Zhang}, \bibinfo{person}{Yiqun Liu},
  \bibinfo{person}{Shaoping Ma}, {and} \bibinfo{person}{Qi Tian}.}
  \bibinfo{year}{2018}\natexlab{}.
\newblock \showarticletitle{Relevance estimation with multiple information
  sources on search engine result pages}. In
  \bibinfo{booktitle}{\emph{Proceedings of the 27th ACM International
  Conference on Information and Knowledge Management}}.
  \bibinfo{pages}{627--636}.
\newblock


\bibitem[Zheng et~al\mbox{.}(2018)]%
        {zheng2018sogou}
\bibfield{author}{\bibinfo{person}{Yukun Zheng}, \bibinfo{person}{Zhen Fan},
  \bibinfo{person}{Yiqun Liu}, \bibinfo{person}{Cheng Luo},
  \bibinfo{person}{Min Zhang}, {and} \bibinfo{person}{Shaoping Ma}.}
  \bibinfo{year}{2018}\natexlab{}.
\newblock \showarticletitle{Sogou-qcl: A new dataset with click relevance
  label}. In \bibinfo{booktitle}{\emph{The 41st International ACM SIGIR
  Conference on Research \& Development in Information Retrieval}}.
  \bibinfo{pages}{1117--1120}.
\newblock


\bibitem[Ziegler et~al\mbox{.}(2005)]%
        {ziegler2005improving}
\bibfield{author}{\bibinfo{person}{Cai-Nicolas Ziegler},
  \bibinfo{person}{Sean~M McNee}, \bibinfo{person}{Joseph~A Konstan}, {and}
  \bibinfo{person}{Georg Lausen}.} \bibinfo{year}{2005}\natexlab{}.
\newblock \showarticletitle{Improving recommendation lists through topic
  diversification}. In \bibinfo{booktitle}{\emph{Proceedings of the 14th
  international conference on World Wide Web}}. \bibinfo{pages}{22--32}.
\newblock


\end{thebibliography}
